\documentclass[aps,pre,showpacs,twocolumn,floatfix,superscriptaddress]{revtex4}
\usepackage{graphicx}
\usepackage{amsmath}
\usepackage{dsfont}

\begin{document}

\title{Self-propelled rods exhibit a novel phase-separated state characterized by the presence of active stresses and the ejection of polar clusters}
%% Giant aggregates and absence of global orientational order in self-propelled rods} 
% \title{Self-propelled rods exhibit a novel type of phase separation and no orientational order} 
% \title{The large-scale properties of self-propelled rods are characterized by the formation of aggregates  and the absence of orientational order} 

\date{\today}

\author{Sebastian Weitz}
\affiliation{Zentrum f\"ur Informationsdienste und Hochleistungsrechnen, Technische Universit\"at Dresden, Zellescher Weg 12, 01069 Dresden, Germany}
\author{Andreas Deutsch}
\affiliation{Zentrum f\"ur Informationsdienste und Hochleistungsrechnen, Technische Universit\"at Dresden, Zellescher Weg 12, 01069 Dresden, Germany}
\author{Fernando Peruani}
\affiliation{Universit{\'e} Nice Sophia Antipolis, Laboratoire J.A. Dieudonn{\'e}, UMR 7351  CNRS, Parc Valrose, F-06108 Nice Cedex 02, France}

\begin{abstract}
We study collections of self-propelled rods (SPR) moving in two dimensions for packing fractions less than or equal to 0.3. 
We find that in the thermodynamical limit the SPR undergo a phase transition between a disordered gas and 
a novel phase-separated system state. Interestingly, (global) orientational order patterns -- contrary to what has been suggested -- vanish in this limit. 
In the found novel state, the SPR self-organize into a  highly dynamical,  high-density, compact region - which we call {\it aggregate} - which is surrounded by a disordered gas. 
Active stresses  build inside  aggregates as result of the combined effect of local orientational order 
 and active forces.  
This leads to the most distinctive feature of these aggregates: constant ejection of polar clusters of SPR. 
This novel phase-separated state represents a  
 novel state of matter characterized by large fluctuations in volume and shape, related to mass ejection, and exhibits  positional as well as orientational local order. 
SPR systems display new physics unseen in other active matter systems. 
\end{abstract}


%% Some pacs of interest. 

% 05.65.+b -- self-organized patterns
% 47.54.-r -- pattern formation
% 87.18.Gh -- Collective behavior of moving cells
% 87.18.Hf -- Spatiotemporal in cell populations
% 87.18.Ed -- cell aggregation
% 05.70.Ln -- non-equilibrim & irreversible physics

\maketitle

\section{Introduction}

Self-organized patterns of self-propelled entities -- from animals to synthetic active particles -- are often believed to be the result 
of a velocity alignment mechanism that regulates the interaction among the moving entities~\cite{vicsek2012,marchetti2013,romensky2014}. 
Such a velocity alignment mechanism can result, for instance, from hydrodynamic interactions~\cite{childress1981,sokolov2007,baskaran2009}, 
electrostatic forces~\cite{bricard2013}, 
moving molecular motors in arrays of microtubules~\cite{bennaim2006,aranson2006},
or from  inelastic bouncing in driven granular particles~\cite{grossman2008,deseigne2010,weber2013}.
There is, in addition, another simple and rather general way of producing a (velocity) alignment mechanism~\cite{peruani2006}:  the combined effect of  steric interactions among elongated objects and self-propulsion  in a dissipative medium, which has been shown to lead to interesting collective phenomena~\cite{peruani2006,yang2010,baskaran2008}.  
%
%
%% If self-propelled elongated rods (SPR) are submerged in a dissipative medium, steric interactions produce forces and torques that lead to an effective velocity alignment mechanism and 
%% collective effects emerge~\cite{peruani2006,yang2010,baskaran2008}. 
%
This mechanism is at work in a broad range of active systems: gliding bacteria~\cite{peruani2012,starruss2012}, driven granular rods~\cite{kudrolli2008,kudrolli2010}, chemically-driven rods~\cite{paxton2004,mano2005}, and 
it has been recently argued that also -- neglecting hydrodynamic effects over steric effects -- in swimming bacteria~\cite{wensink2012,dunkel2013,zhang2010} and motility assays~\cite{schaller2010,sumino2012}. 
%%%%%%%%%%%%%%%%%%%%
\begin{figure}
\begin{center}
\resizebox{\columnwidth}{!} {\includegraphics{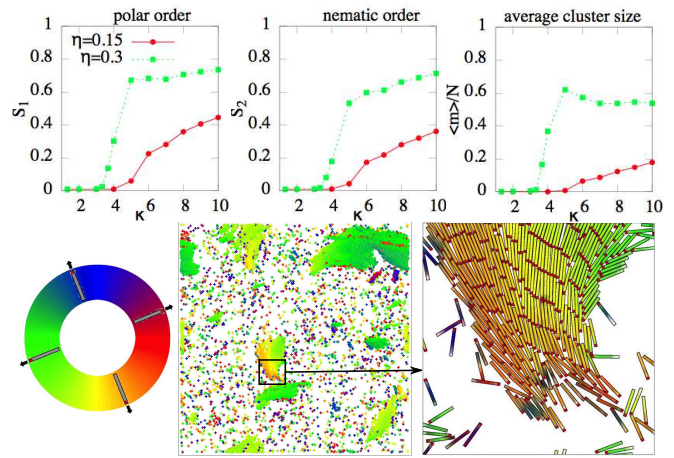}}
%%\resizebox{15cm}{!}{\rotatebox{0}{\includegraphics{figure1.pdf}}}
\caption{Transition from disorder to order and phase separation at small system sizes. Top, from left to right: polar order parameter $S_1$, nematic order parameter $S_2$, and average cluster size $\langle m \rangle$ over system size $N$, respectively, as function of the particle aspect ratio $\kappa$ for various packing fractions $\eta$. $N=10000$. Bottom: A simulation snapshot for $\kappa=10$ and $\eta=0.3$, with $S_1=0.5$, $S_2=0.26$, $\langle m \rangle/N=0.14$. Notice that clusters are polar. The orientation of rods is color coded as indicated in the bottom left panel. This convention is also used in Figs.~\ref{fig:band} and~\ref{fig:aggregate}. 
 }
\label{fig:PT}
\end{center}
\end{figure}
%%%%%%%%%%%%%%%%%%%%%%%%

Here, we focus on the large-scale physical properties of collections of self-propelled rods (SPR) for packing fractions $\eta$ less than or equal to $0.3$. 
We find that the combined effect of steric repulsive forces and active forces lead to a complex interplay between local orientational order and active stresses. 
As result of such interplay, SPR exhibit, for large enough system sizes, new physics unseen in other active matter systems. 
The novel phenomena reported here passed unnoticed in previous SPR studies that were performed with either small system sizes or lack a finite size study~\cite{peruani2006,yang2010,baskaran2008,dunkel2013,wensink2012b,abkenar2013,mccandlish2012}. 
In particular, we provide strong evidence that global orientational order patterns (see Fig.~\ref{fig:PT}), suggested to exist for $\eta \leq 0.3$~\cite{wensink2012,wensink2012b}, vanish in the thermodynamical limit. 
More importantly, we find that SPR undergo a phase transition between a disordered gas and 
a novel phase-separated system state characterized by the presence of active stresses and the ejection of polar clusters of SPR. 

The paper is organized as follows. We start out by introducing the self-propelled rod model in Sec.~\ref{sec:model}. 
The statistical features of the phase-transition in finite systems is studied in Sec.~\ref{sec:phase-transition}.  
A finite-size analysis is presented in Sec.~\ref{sec:FSS}.  
The remarkable dynamics displayed by aggregates in large systems is analyzed in Sec.~\ref{sec:aggregate}.   
In Sec.~\ref{sec:discussion} we discuss  the obtained results.

\section{Model definition}  \label{sec:model}

%% {\it Model definition.--}
%
Our model consists of  $N$ SPR moving in a two-dimensional space of linear size $L$ with periodic boundary conditions. 
Each rod is driven by an active stress/force $F$ that is applied along the long axis of the particle. Interactions among rods are modeled through 
a repulsive potential, which we denote, for the $i$-th particle, by  $U_i$. 
The substrate  where the rods move acts as a momentum sink.  
There are three friction drag coefficients,  $\zeta_{\parallel}$, $\zeta_{\perp}$, and $\zeta_{\theta}$, 
which correspond to the drags experienced by the rods as the rod moves along the long axis, perpendicular to it, or as it rotates, respectively.
%% ~\cite{commentFriction}.  
%
In the over-damped limit, the equations of motion of the $i$-th 
rod are given, as in~\cite{peruani2006}, by:
\begin{eqnarray}
\label{eq:evol_x}
\dot{\mathbf{x}}_i &=&  \boldsymbol{\mu}  \left[ -\boldsymbol{\nabla}U_i +  F \mathbf{V}(\theta_i)+  \boldsymbol{\sigma}_i(t) \right]  \\
\label{eq:evol_theta}
\dot{\theta}_i       &=&    \frac{1}{\zeta_{\theta}}  \left[  - \frac{\partial U_i}{\partial \theta_i} +   \xi_{i}(t) \right]  \, , 
\end{eqnarray}
where the dot denotes a temporal derivative, $\mathbf{x}_i$ corresponds to the position of the center of mass and $\theta_i$ the orientation of the long axis of the rod. 
%
%The term $U_i$ is an interaction potential and $F$ is the self-propelling force. 
%
In Eq.~(\ref{eq:evol_x}), $\boldsymbol{\mu}$ is the mobility tensor defined as  $\boldsymbol{\mu} = \zeta_{\parallel}^{-1} \mathbf{V}(\theta_i) \mathbf{V}(\theta_i) + \zeta_{\perp}^{-1} \mathbf{V}_{\perp} (\theta_i) \mathbf{V}_{\perp}(\theta_i)$, 
with $\mathbf{V}(\theta)\equiv (\cos(\theta),\sin(\theta))$ and $\mathbf{V}_{\perp}(\theta)$ such that  $\mathbf{V}(\theta).\mathbf{V}_{\perp}(\theta)=0$.
We use $\zeta_\parallel=10$, $\zeta_\perp=25$, $\zeta_\theta=2$, and $F=0.4$, which corresponds to an active speed $v_0 = F/\zeta_{\parallel}=0.04$.  
Other friction coefficient values, as well as drag friction models, have been also tested as detailed in~\cite{commentFriction}, obtaining the same qualitative results. 
For details about how to compute drag friction coefficients we refer the reader to~\cite{doi1986,levine2004}. 
%
% and $F$ is the magnitude of the active stress, which defines the active speed $v_0 = F/\zeta_{\parallel}$. 
%
The temporal evolution of the orientation of the rod, given by Eq.~(\ref{eq:evol_theta}), results from the torque $-\frac{\partial U_i}{\partial \theta_i}$ generated by the interactions and no active torque is present. 
Eqs. (\ref{eq:evol_x}) and~(\ref{eq:evol_theta}) are subject to fluctuations through 
the terms $\boldsymbol{\sigma}_i(t)$ and $\xi_{i}(t)$, which correspond to  delta-correlated vectorial and scalar noise, respectively. 
%
%% This suggests that fluctuations are relatively more important in Eq.~(\ref{eq:evol_theta}) than in Eq.~(\ref{eq:evol_x}). 
%
If these fluctuations are of thermal origin, it can be shown that the 
``passive" diffusion coefficient $D_p$ resulting from $\boldsymbol{\sigma_i}$  (i.e. the diffusion for $F=0$)  
is negligible compared to the active diffusion coefficient given by $D_a$ (corresponding to $F>0$ and $\boldsymbol{\sigma_i}=\boldsymbol{0}$); for details see Appendix~\ref{sec:DemoDaSmallerThanDp}.
Since $D_p/D_a \ll 1$, for simplicity we neglect $\boldsymbol{\sigma_i}$ and specify in  Eq.~(\ref{eq:evol_theta})  $\langle \xi_{i}(t) \rangle = 0$ 
and $\langle \xi_{i}(t) \xi_{j}(t') \rangle = 2 D_{\theta} \delta_{i,j} \delta(t-t')$, with $D_{\theta}=2.52\times10^{-2}$. 
%
%
% = v_0^2/(2D_{\theta})$
%Furthermore,  we assume that the ``passive" diffusion coefficient $D_p$ resulting from $\boldsymbol{\sigma_i}$ is negligible compared to the 
%active diffusion coefficient given by $D_a = v_0^2/(2D_{\theta})$, i.e. $D_a >> D_p$, and neglect the term $\boldsymbol{\sigma_i}$.  
%%
%In summary, we neglect fluctuations in Eq.~(\ref{eq:evol_x}) and consider only the noise in Eq.~(\ref{eq:evol_theta}) characterized by $\langle \xi_{i}(t) \rangle = 0$ 
%and $\langle \xi_{i}(t) \xi_{j}(t') \rangle = A_{\theta} \delta_{i,j} \delta(t-t')$, with $A_{\theta}=5\times10^{-2}$. 
%%% $D_\theta=0.0063$. 
%
The interactions among the rods are modeled by a soft-core potential that penalizes particle overlapping. For the $i$-th rod, the potential takes the form:   
$U_i = U(\mathbf{x}_i, \theta_i) = \sum_{j=1;j \neq i}^{N} u_{i,j}$ ,
% \begin{eqnarray}
% \label{eq:potential}
% U(\mathbf{x}_i, \theta_i) = \sum_{j=1;j \neq i}^{N} u_{i,j} \, ,
% \end{eqnarray}
%
where $u_{i,j}$ denotes the repulsive potential interaction between the $i$-th and $j$-th rod, both of length $\ell$ and width $w$, 
such that  $u_{i,j}=u(\mathbf{x}_i-\mathbf{x}_j, \theta_i-\theta_j)$.
The rods  are represented as  straight chains of $n$ disks of diameter $w$, as implemented in~\cite{wensink2012b,abkenar2013,mccandlish2012}, whose centers are separated at distance $\Delta=w/3$. 
We notice that results obtained with SPR represented by disk-chains are qualitatively identical to those produced with the original SPR model introduced in~\cite{peruani2006} 
if and only if $\Delta \ll 2w$; for more details see~\cite{commentRodDisks}.  
Using this implementation,  $u_{i,j}$ can be  expressed as  $u_{i,j}=\sum_{\alpha, \beta} u_{i,j}^{\alpha, \beta}$, where 
$u_{i,j}^{\alpha, \beta}$ is the potential between disk $\alpha$ of the $i$-th rod and disk $\beta$ of the $j$-th rod, which here we assume to be given by a harmonic repulsive potential:  
$u_{i,j}^{\alpha, \beta}=C_0\left(d^{i,j}_{\alpha, \beta}-w\right)^2$, for $d^{i,j}_{\alpha, \beta}<w$ and zero otherwise, 
where $d^{i,j}_{\alpha, \beta}$ is the distance between the centers of the disks and $C_0=200$. 
\begin{figure}
\begin{center}
\resizebox{\columnwidth}{!} {\includegraphics{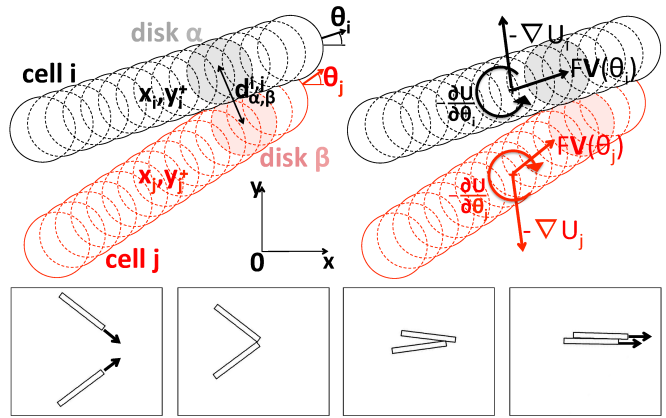}}
\caption{Interactions in the SPR model.
Top: Sketch of two interacting rods in the SPR model.
Forces and torques resulting from the interaction are shown in the top-right panel. See text for explanations.
Bottom) Chronological snapshots of a collision between two rods. Notice that even though the interaction is exclusively repulsive, it leads  to an effective velocity alignment (an   effective attraction). A movie of this interaction is provided in~\cite{commentSI}.
}
\label{fig-model}
\end{center}
\end{figure}

Fig.~\ref{fig-model}  illustrates the implementation of the interaction between two rods. 
Rod $i$ is propelled forward by its active force $F \mathbf{V}(\theta_i)$, while it is pushed away by rod $j$ through the interaction force $-\mathbf{\nabla}U_i$. The interaction with rod $j$ also leads to a torque, given by $-\frac{\partial U_i}{\partial \theta_i}$. This torque, together with the over-damped dynamics, lead to an effective alignment of the velocity of the rods as shown in the bottom panels of Fig.~\ref{fig-model} and previously described in~\cite{peruani2006}. The snapshots show, in chronological order, a collision event between two rods: The rods are moving in different directions (indicated by the arrows) before the collision (first snapshot). They start colliding (second snapshot). The steric forces and torques resulting from the interaction lead to an  effective alignment (third snapshot). As result, both rods end up moving in roughly the same direction and stay close to each other (last snapshot) - without requiring attractive force. 
Notice that interestingly the described alignment process is similar to the one predicted to occur using kinetic theory in model for ordering of microtubules mediated by molecular motors in a planar geometry~\cite{bennaim2006,aranson2006}.

%A binary ``collision" between rods, which is not an instantaneous event 
%given that the equations of motions are  over-damped, see Eq.~(\ref{eq:evol_x}) and~(\ref{eq:evol_theta}), lead to an effective alignment of the velocity of the rods 
%as indicated by the bottom panels in Fig.~\ref{fig-model}. 

%
% \begin{eqnarray}
% \label{eq:harmonic}
% u_{i,j}^{\alpha, \beta}=
% \begin{cases} 
% \frac{k}{2}\left(d^{i,j}_{\alpha, \beta}-w\right)^2 &\mbox{if } d^{i,j}_{\alpha, \beta}<w \\
% 0 & \mbox{otherwise} \, ,  
% \end{cases}
% \end{eqnarray}
% %
% where $d^{i,j}_{\alpha, \beta}$ is the distance between the disks~\cite{commentModels}. The model parameter values are given in~\cite{commentParameters}.

%% \bibitem{commentParameters} The shown numerical data  correspond to the following parameter values:
% $\Delta=w/3$ (equivalently, $n=3\kappa$), 
%$D_\theta=0.0063$, 
%$\zeta_\parallel=10$, 
%$\zeta_\perp=25$, 
%$\zeta_\theta=2$, 
%$v_0=0.04$, 
%$k=400$, 
%$a=l\,w=0.1$, and 
%% $\Delta t=0.25\kappa^{-1}$. Simulations are implemented using a Euler-Maruyama scheme.
 
\section{Transition from disorder to order and phase separation in small finite systems} 
\label{sec:phase-transition}
%% {\it  Results.--}
%
We explore the parameter space of the model using parallelized numerical simulations implementing an Euler-Maruyama scheme.
As control parameter, we use the rod aspect ratio $\kappa = \ell /w$ and keep constant the particle area $a=\ell \times w=0.1$ as well as all the other parameters.     
We perform the analysis for  several values of $\eta$. Notice that $\eta$ can be also used as control parameter. 
The macroscopic patterns are characterized by their level of orientational order through 
\begin{eqnarray}
\label{eq:op_order}
 S_q   = \langle S_q(t) \rangle_t = \langle | \langle \exp(\imath \,q\,\theta_i(t)) \rangle_{i} | \rangle_t \, ,
\end{eqnarray}
where the averages run over the number of particles and time. Polar order corresponds to $q=1$ and  nematic order to $q=2$. 
Phase separation is monitored by looking at the ratio between average cluster size $\langle m \rangle$ and system size $N$, where $\langle m \rangle =  \sum_m m\, p(m)$ with $p(m)$ the (weighted steady state) cluster size distribution. $p(m)$ is defined as the time average (after an initial transient) of the instantaneous cluster size distribution 
\begin{eqnarray}
p(m,t)=\frac{m~n_m(t)}{N},
\end{eqnarray}
where $n_m(t)$ is the number of clusters of mass $m$ present in the system at time $t$.
Notice that the normalization of this distribution is ensured since $N=\sum_{i=1}^N m~n_m(t)$. 
%
%We expect $n_m(t)$ after a transient to fluctuate around a steady state value. We define the cluster size distribution $p(m)$ as:
%\begin{eqnarray}
%p(m)=\frac{m~\langle n_m(t) \rangle_t }{N} \,
%\end{eqnarray}
%where the average is taken after the above mentioned transient. 
%
Clusters are collections of interconnected rods, where any two rods are considered as connected if they are separated by a distance equal to or smaller than $2w$, 
implemented as the minimum distance between the centers of the disks that form each rod. 

Varying the aspect ratio  $\kappa$, while keeping fixed  the packing fraction $\eta = a \,N/L^2$,   
we observe a clear phase transition with $S_1$, $S_2$, and $\langle m \rangle/N$ taking off above a critical $\kappa_c$ value as shown in Fig.~\ref{fig:PT}. We notice that $\kappa_c$ decreases when $\eta$ is increased.
Below $\kappa_c$, we observe a gas phase characterized by the absence of orientational order and an exponential cluster size distribution 
such that  $\langle m \rangle/N \sim \mathcal{O}(N^{-1})$. 
For $\kappa>\kappa_c$ the system undergoes  a symmetry breaking as observed previously in~\cite{baskaran2008,wensink2012b,abkenar2013}. 
The emerging order is  polar as evidenced by the behavior of $S_1$. 
We recall that in the presence of polar order, $S_2$  is slaved to $S_1$. 
%In all configurations, we observe that the behavior of $S_2$ follows closely the behavior of  $S_1$. This is because (in finite systems) we observe polar order. 
%
The behavior of $\langle m \rangle/N$, right panel in Fig.~\ref{fig:PT}, indicates that  the system starts to spontaneously self-segregate for $\kappa>\kappa_c$. 
Here, we find that the onsets of orientational order and phase separation coincide and share the same critical point, as predicted using a simple 
kinetic model for the clustering process~\cite{peruani2013}: due to the effective velocity alignment large polar clusters emerge, which in turn lead to macroscopic polar order. 
Notice that in an equilibrium system of (hard) rods (i.e. $F=0$) for $\eta \leq 0.3$ and $1 \leq \kappa \leq10$, according to De las Heras et al.~\cite{delasheras2013}, 
we should observe only an homogeneous disordered phase for this range of parameters.  
This indicates that the observed phase transition requires $F>0$, i.e. the active motion of the rods. 

\section{Finite size study: absence of global order in the thermodynamical limit}
\label{sec:FSS}

%%%%%%%%%%%%%%%%%%%%
\begin{figure} 
\begin{center}
%\resizebox{\columnwidth}{!} {\includegraphics{Fig2.eps}}
%%\resizebox{15cm}{!}{\rotatebox{0}{\includegraphics{figure1.pdf}}}
\resizebox{\columnwidth}{!} {\includegraphics{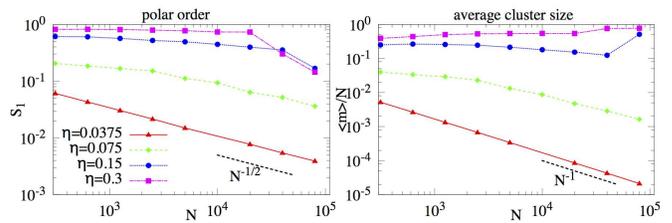}}
\caption{Finite size scaling of the polar order parameter $S_1$ and average cluster size with respect to system size $\langle m \rangle/N$ for aspect ratio $\kappa = 10$ and several packing fractions $\eta$. 
Notice that, for $\eta \geq 0.15$, the system becomes disordered as the system size $N$ is increased, while remaining phase-separated.  
 }
\label{fig:FSS}
\end{center}
\end{figure}
%%%%%%%%%%%%%%%%%%%%%%%%

We performed a finite size study, by increasing simultaneously $N$ and $L$ while keeping the packing fraction $\eta$ and all other parameters constant. Fig.~\ref{fig:FSS} shows the scaling of the (global) polar order parameter $S_1$ and average cluster size with respect to system size $\langle m \rangle/N$ for aspect ratio $\kappa = 10$ and several packing fractions $\eta$. 
At low $\eta$ values, i.e. for $\eta \leq 0.075$, we are in the situation $\kappa < \kappa_c$ (we recall that the critical $\kappa_c$ value depends on $\eta$). We showed in Sec.~\ref{sec:phase-transition} that  for $\kappa < \kappa_c$ the system is not phase-separated and does not exhibit orientational order. As expected, the scaling of $S_1$ with $N$ shows that  $S_1 \propto N^{-\alpha}$, with $\alpha = 1/2$, which means that the system is fully disordered. In addition, we observe that  $\frac{<m>}{N} \propto N^{-\beta}$, with $\beta = 1$, which indicates that there is a well-defined characteristic cluster size for the system (that is independent of $N$) and consequently the system is not phase-separated.

At large $\eta$ values, i.e. for $\eta \geq 0.15$ and $\kappa > \kappa_c$,  we observe phase separation and (global) polar order for small finite systems ($N=10000$), Sec.~\ref{sec:phase-transition}. The finite size study shows that, for $\kappa > \kappa_c$, $\frac{<m>}{N}$ does not decrease (asymptotically) with $N$. Moreover, for large values of $N$, we even observe an increase. This indicates that $\langle m \rangle$ is at least proportional to $N$, which implies that the system is phase-separated in the thermodynamical limit as well as in finite systems.
At the level of the orientational order  parameter $S_1$, we observe an abrupt change in scaling of $S_1$ with $N$. 
For $N<N_*$ (e.g., $N_*\sim 20000$) for $\eta=0.3$ and $\kappa=10$, $S_1$ is high {\it i.e.} the system displays  global polar order . 
On the other hand,  for $N>N_*$, while the system remains phase-separated, $S_1$ sharply decreases with $N$. 
In short, the finite size study reveals that, although phase separation does take place in the thermodynamical limit as well as in finite systems, the phase transition to an orientationally ordered phase, described in Sec.~\ref{sec:phase-transition}, is observed only for small finite systems. 
%
%The (global) polar order parameter $S_1$  decreases with system size $N$ which suggests that 
Global order patterns are not present in the thermodynamical limit, with the phase transition occurring, in this limit, between a disordered gas and a phase-separated state with no global orientational order. 
 %
%In contrast, $\langle m \rangle/N$ does not decrease (asymptotically) with $N$, which indicates that phase separation does take place in the thermodynamical limit as well as in finite systems.
%

%%%%%%%%%%%%%%%%%%%%
\begin{figure} [h!]
\begin{center}
%\resizebox{\columnwidth}{!} {\includegraphics{Fig3.eps}}
%%\resizebox{15cm}{!}{\rotatebox{0}{\includegraphics{figure1.pdf}}}
%\includegraphics[width=\columnwidth]{SM/S1-histograms.pdf}
%\includegraphics[width=\columnwidth]{SM/epot-dynamics.pdf}
\resizebox{\columnwidth}{!} {\includegraphics{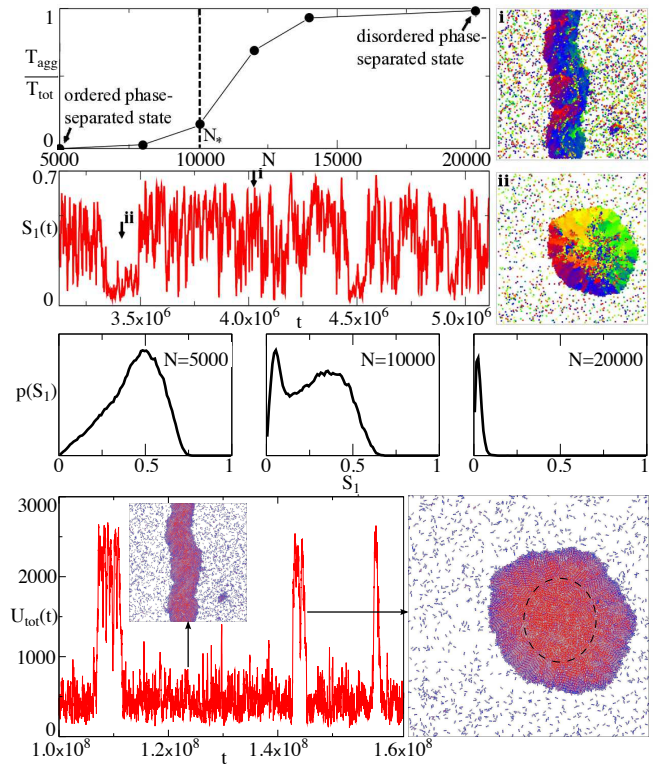}}
\caption{
The orientationally ordered phase-separated state observed for $\kappa>\kappa_c$ in small systems  ($N\ll N_*$) becomes instable when $N$ is increased. In very big systems ($N\gg N_*$) we observe only aggregates (which correspond to an orientationally disordered phase-separated state).
Top left: Evolution of  $T_{agg}/T_{tot}$ with $N$, where  $T_{agg}$ is the time the system spent in the aggregate phase and $T_{tot}$ -- here, $T_{tot}=10^7$ --  is the total simulation time.
Second line, left: the polar order $S_1$ as function of time for $N\sim N_*$ -- here, $N_*\sim 10^4$.
Large values of $S_1$ correspond to the system being highly ordered, typically due to the formation of a highly ordered band (panel i), 
while low values of   $S_1$ correspond to the formation of an aggregate (panel ii).  
Third line: Histograms of the polar order parameter $p(S_1)$ for $N=5000$, $N=10000\sim N_*$ and $N=20000$.
Bottom panel: Total elastic energy $U_{tot}=\sum_{i=1}^{N}U_i$ of the system as function of time $t$, for $N\sim N_*$. 
High values of $U_{tot}$ correspond to the presence of an aggregate (inset on the left panel), while 
low values of $U_{tot}$ are related to the presence of polar structures such as bands. 
On both snapshots, the color of each rod indicates its interaction potential $U_i$: blue (red) color indicates small (large) $U_i$ values.
The dashed black circle in the snapshot corresponding to an aggregate provides an idea of the boundary shell of the aggregate. 
For movies, see~\cite{commentSI}. 
Simulations correspond to $\kappa = 4$ and $\eta = 0.3$.
}
\label{fig:band}
\end{center}
\end{figure}
%%%%%%%%%%%%%%%%%%%%%%%%

%
The reason for observing non-vanishing global order in small systems is the presence of few giant polar clusters as illustrated by the simulation snapshot in Fig.~\ref{fig:PT}.
%that these system sizes (think of $L$) are smaller than the correlation length of the orientations of the SPR, as indicated by the slow decay of $S_1$ with $N$ for $N<N_*$, together with the large value displayed by this order parameter. 
%
Such giant polar clusters can become so big and elongated that they can can even percolate the system, as shown in panel i) of Fig.~\ref{fig:band}.
We refer to such  polar percolating structures as bands. Inside bands, rods are densely packed, point into the same direction, and exhibit  positional order. 
Notice that these bands are distinct from the bands observed in the Vicsek model, which are elongated in the direction orthogonal to the moving direction of the particles~\cite{gregoire2004}. 
The observed polar bands are also fundamentally different from those observed in models of idealized SPR, where the point-like self-propelled particles 
form nematic bands, inside which 50\% of the particles move in one direction and 50\% in the opposite one~\cite{ginelli2010}. 
More importantly, our finite size study indicates that the polar patterns observed in SPR are a finite size effect that disappear for large enough systems. 
In short, several of the phases reported for $\eta \leq 0.3$  in previous SPR works~\cite{wensink2012, wensink2012b} 
such as the so-called swarming phase and the bio-turbulence phase vanish in the thermodynamical limit. 

The abrupt change in scaling of $S_1$ with $N$ in Fig.~\ref{fig:FSS} suggests that above the crossover system size $N_*$ the polar structures are no longer stable.
%While the system remains phase-separated, the polar order observed in the situation $N<N_*$ (leading there to giant polar moving clusters and bands) is no longer stable for $N>N_*$ because the system size ($L$) is larger than the correlation length of the SPR. This situation corresponds to the existence of a giant  {\it aggregate} that exhibits no net displacement (i.e. no polar order), which is surrounded by a dilute gas phase.
%
Arguably, the decay in $S_1$ with $N$ is due to the fact that rods inside polar clusters are densely packed and hold fixed positions, not being 
able to exchange neighbors in contrast to other active systems~\cite{vicsek1995,ginelli2010,marchetti2012b,abkenar2013,gregoire2004}. 
In the co-moving frame that moves with the cluster, we have a two-dimensional system of particles interacting locally and subject to fluctuations. 
Assuming that in this scenario we can apply the Mermin-Wagner theorem~\cite{mermin1966}, long-range order is not possible and for sufficiently big clusters defects in the orientation of the rods 
should emerge. If such defects are present in a cluster/band, the velocity field of the polar structure will be necessarily unstable (see also Appendix~\ref{sec:polar-order-and-clusters}). 
%% , see~\cite{commentSI}. 
%
The instability of polar structures is evident by looking at the behavior of bands with $N$, Fig.~\ref{fig:band}.
The top panel of this figure shows the finite size scaling of $T_{agg}/T_{tot}$, {\it i.e.} the total time $T_{agg}$ the system spends in the aggregate phase with respect to the total simulation time $T_{tot}$. Note that the computation of  $T_{agg}$ implies looking for all events where an aggregate emerged in the system, accumulating the time each aggregate lived. 
We observe that $T_{agg}$  increases with $N$, in such a way that $T_{agg}/T_{tot} \to 1$as $N\to\infty$.  This means that the probability of observing the system in an aggregate phase also increases with $N$. 
For small system sizes $N \ll N_*$ we observe moving clusters and bands. Large polar structures such as bands form, remain in the system for  quite some time, and then quickly break and reform, typically adopting a new orientation. The corresponding histogram of global polar order -- $p(S_1)$ -- in Fig.~\ref{fig:band} ($N=5000$) is unimodal with a peak at large values of $S_1$.
As $N \to N_*$,  bands  survive for relatively short periods of time, and quickly bend and break. Interestingly, at such large system sizes other macroscopic structures start to frequently emerge.  
These new macroscopic structures -- which we refer to as {\it aggregates} --  are formed by polar clusters of rods that exert stresses on each other and exhibit vanishing polar order, see panel ii) of Fig.~\ref{fig:band}. 
%
% with rods a
% exhibit a large external ring, where rods point inwards, and a core that contains locally ordered cluster of rods. 
%
In summary, for $N \sim N_*$, the system continuously transitions between highly ordered phases -- e.g. phases with either a few giant polar clusters or a band -- and aggregates, as illustrated in Fig.~\ref{fig:band}.  The corresponding histogram of $S_1$ ($N=10000$) is bimodal with a peak at large values of $S_1$, corresponding to the polar structures, and another peak at very small values of $S_1$, corresponding to aggregates. 
As the system size is increased further, i.e. for $N \gg N_*$, we observe that the corresponding histogram of $S_1$ ($N=20000$) becomes again unimodal, but the peak is now at very small values of $S_1$, and corresponds to the presence of aggregates.
In short, bands and polar phases disappear in the thermodynamical limit, while the aggregate phase survives (see also the phase diagrams in Appendix~\ref{phase-diagrams}).  
The dynamics of aggregates for large systems size is studied in details in Sec.~\ref{sec:aggregate}.

The continuous transitions between aggregates and bands (or highly ordered phases) for $N \sim N_*$ 
 results from the competition between elastic energy and the impossibility of the system to sustain long-range polar order. 
For not too large system sizes, {\it i.e.} for $N\sim N_*$, the shape of the aggregates is roughly circular (Fig.~\ref{fig:band}, panel ii)) 
and  at the center of the aggregate we find one single topological defect: i.e. at the mesoscale, at the center of the aggregate we cannot define an average orientation for the rods. 
Due to the active forces,  at the center of the aggregate rods are strongly compressed, which implies that the  potentials $U_i$ adopts 
high values (see Fig.~\ref{fig:band}, bottom row). 
This implies that when one of these aggregates is formed, the total elastic energy of the system $U_{tot}=\sum_{i=1}^{N}U_i$ increases. 
On the contrary, in large polar structures such as bands, rods are roughly parallel to each other and therefore are much less compressed by their neighbors, and   
the  total elastic energy is low. 
This is evident on the bottom left panel of Fig.~\ref{fig:band}. 
The dynamics at $N \sim N_*$ can be summarized as follows. Large polar clusters form and eventually a band emerges, 
but since the system is too big for the band to remain stable, at some point the band breaks. 
The collapse of the band gives rise to the formation of new giant polar clusters which eventually collide head on leading to a large aggregate: a process reminiscent of a traffic jam. 
The formation of the aggregate leads to a sharp increase of the total elastic energy. Let us recall that forces and torques act in such a way that they tend to minimize $U_i$. 
In short, the system relaxes by destroying the new formed aggregate, which give rise to the formation of new polar clusters and the cycle starts again. 
In larger system sizes, i.e. for  $N \gg N_*$, aggregates are more complex. This is addressed in the next section.

%%%%%%%%%%%%%%%%%%%%
\begin{figure} [h!]
\begin{center}
%\resizebox{\columnwidth}{!} {\includegraphics{Fig4.eps}}
%\resizebox{\columnwidth}{!} {\includegraphics{SM/smectic.pdf}}
\resizebox{\columnwidth}{!} {\includegraphics{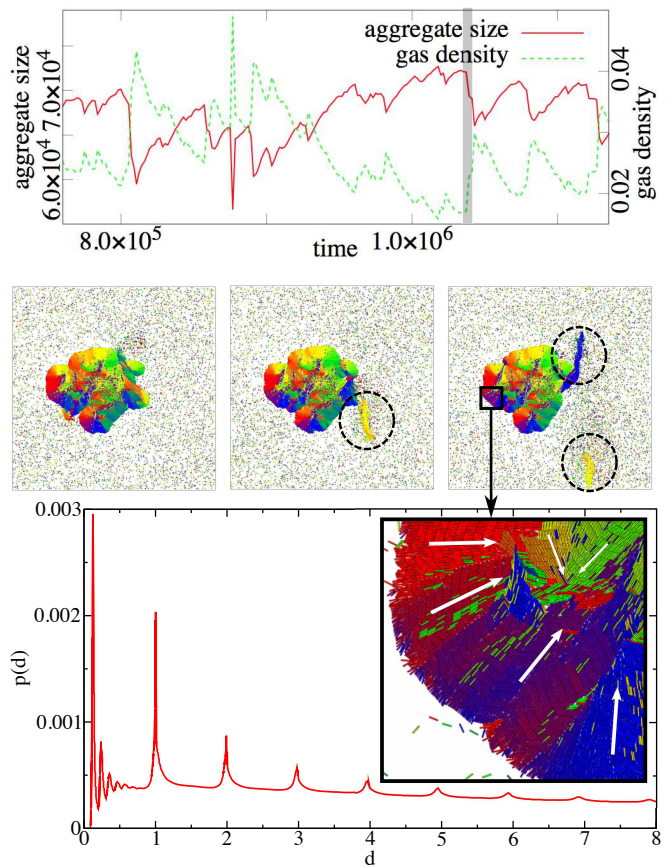}}
\caption{Dynamics of an aggregate. 
Top panel: aggregate size and gas density as function of time. 
The aggregate boundary exhibits large fluctuations due to the emergence of orientational defects that lead to the detachment of large polar clusters from the aggregate. 
Second line: The three panels display in chronological order one of these events. The corresponding time window is indicated by the vertical grey area in the top left panel.  Dashed circles indicate the detachment of polar clusters. The inset shows that topological defect.
Bottom panel: Orientational and positional order within in the aggregates. Left:
$p(d)$ is the probability density that the distance between the centers of two rods is $d$ (in units of $\ell$) . We observe peaks at multiples of the rod length $\ell$ (the smaller peaks between $0$ and $1$ correspond to $w$). This is the fingerprint of positional (smectic-like) order: the rods are arranged as on a lattice. 
Right: Snapshot of an aggregate.
Inset: Zoom on a subdomain of the aggregate. 
The dashed black circle in the snapshot provides a visual estimate of what we call the core of the aggregate. Colors encode rod orientation to show the domains of similarly orientated rods. The arrows indicate the local orientation of the rods. 
Simulations of the aggregate dynamics correspond to $N=80000$, $\eta=0.15$, and $\kappa=7$ ($N=10000$, $\eta=0.3$, and $\kappa=4$ for the aggregate in the bottom panel). For movies, see~\cite{commentSI}. 
%Simulations for the smectic order diagram correspond to $N = 10000$, $\eta= 0.15$ and $\kappa=10$.
% 
 }
\label{fig:aggregate}
\end{center}
\end{figure}
%%%%%%%%%%%%%%%%%%%%%%%%

\section{Aggregate dynamics}
\label{sec:aggregate}

Inside aggregates, the competition between active forces and local polar alignment leads to new physics unseen 
in other active systems.  
This is particularly evident for very large system sizes, i.e. for $N\gg N_*$, 
that is when aggregates are big enough to exhibit multiple topological defects of the local orientation of the rods.

Let us recall that aggregates are formed by polar clusters of rods that are trapped inside these structures.
The rods inside the aggregate do not only exhibit local orientational order, but also local positional order as indicated by the peaks exhibited by 
the probability density $p(d)$ of finding the center of a rod at a distance $d$ of the center of another rod (see bottom panel of Fig.~\ref{fig:aggregate}). 
Topological defects are areas where, at the mesoscale, as mentioned above, we cannot define an average orientation of the rods as illustrated in the bottom inset of Fig.~\ref{fig:aggregate} (areas where the arrows meet). 
In such areas, due to the active forces, rods are strongly compressed by the active push of all surrounding SPR (see inset in Fig.~\ref{fig:aggregate}). 
Since the compression is due to the presence of active forces, we refer to this phenomenon as {\it active stresses}. 
For $N\gg N_*$, we observe the emergence of multiple defects that lead to an increase of the elastic energy and the build-up of stresses. Notice that more topological defects imply larger values of the elastic energy. 
There are two clear consequence of the presence of multiple topological defects. On the one hand, aggregates are no longer roundish but rather irregular as illustrated in Fig.~\ref{fig:aggregate}. 
On the other hand, now the system can relax the elastic energy by reducing the number of topological defects. 
Notice that for $N \sim N_*$, aggregates are relatively small and exhibit one topological defect, 
and thus, the only way to eliminate the topological defects is by destroying the aggregate. 
For $N\gg N_*$, given the presence of multiple topological defects, eliminating one topological defect does not require to eliminate the aggregate.  
As a matter of fact, for very large system sizes, 
the interplay between topological defects and active stresses lead to  large fluctuations of the aggregate boundary and aggregate mass (i.e. aggregate size) as indicated in Fig.~\ref{fig:aggregate}. 
The most distinctive feature of the observed phenomenon is the large fluctuations experienced by the aggregate mass 
correspond to ejections of remarkably large macroscopic polar clusters from the aggregate, that can be as large as 10\% of the system size (i.e. involving more than $10^4$ rods), Fig.~\ref{fig:aggregate}. 
By this process, i.e. the ejection of large polar clusters, the aggregate manages to decrease its elastic energy. 
The ejected polar clusters typically dissolve while moving through the gas phase outside the aggregate, leading to a sudden increase of the gas density,  top panel in Fig.~\ref{fig:aggregate}. This results in a higher absorption rate of SPR by the aggregate that starts again to increase its mass.   
The system dynamics - in the phase-separated state - can be summarized (in a simplified way) as follows:  Aggregates  grow in size and multiple defects emerge inside the aggregate. This results in  active stresses that build up and lead to fluctuations of the aggregate boundary and ejection of huge polar clusters. This implies a reduction of the aggregate size and its elastic energy, and an increase of the gas density at which point the cycle starts again. 

\section{Discussion} \label{sec:discussion}

The ejections of thousands of particles in densely packed and highly ordered clusters -- the most distinctive feature of the described dynamical phase-separated state -- 
requires  the combined effect of an effective alignment mechanism and active pushing (or stresses) acting among the particles. 
The combination of these two elements is not present, to the best of our knowledge, in any other active matter model and is a distinctive property of SPR. 
And even for SPR such effects are only evident for large enough system sizes, i.e. above $N_*$.  
For instance, self-propelled disks (SPD) exhibit active stresses but no alignment among the SPD~\cite{fily2012,buttinoni2013,redner2013}. While in SPD systems phase separation is also observed, 
and high-density objects as aggregates are found, the dynamics of these objects is totally different: SPD aggregates are not formed by polar clusters  
of SPD and  they do not eject huge polar clusters of active particles. 
Moreover, SPD inside the aggregates are contained by a thin ring of SPD pointing inwards, which suggests that 
in the presence of several aggregates, particle exchange occurs via an evaporation-like process, controlled by the angular diffusion coefficient, 
leading to a classical coarsening process~\cite{fily2012,buttinoni2013,redner2013}, which can be described by an effective Cahn-Hilliard equation~\cite{speck2014}.    
In sharp contrast with this scenario, the phase separation of SPR starts with a non-equilibrium ballistic clustering process that leads to the formation of polar clusters~\cite{peruani2013}, 
which in turn collide ballistically until eventually a traffic jam of polar clusters occurs and an aggregate emerges. 
The density instabilities and fluctuations observed with SPR are also remarkably different from what has been reported in systems of point-like active particles with an alignment mechanism 
as in the Vicsek models~\cite{vicsek2012,marchetti2013}, where density fluctuations and phase separation (in the form of bands) require 
the presence of either polar or apolar long-range or quasi-long range order~\cite{dey2012}, 
which are absent in SPR as shown here. 
Finally, though in active particle systems that combine an alignment mechanism and a density-dependent speed~\cite{marchetti2012b,peruani2011} 
 polar clusters, bands, and aggregates are found, as occurs in SPR for small system sizes, the large-scale properties of these systems are radically different 
from what has been reported here for SPR. In such systems, there is no active push among the active particles and while 
topological defects cannot be ruled out, it can be safely stated that they cannot generate active stresses. 
Thus, in such models bands and polar phases might exist even in the thermodynamical limit, and certainly aggregates cannot eject polar clusters. 

In summary, the physics of SPR is remarkably different from the one observed in any other particle system.  
The coupling between active stresses and order -- that implies that topological defects and active stresses are intimately related -- leads to novel phenomena as the here reported phase-separated state characterized by the ejection of polar clusters.

\appendix

\section{Demonstration that $D_p\ll  D_a$}
\label{sec:DemoDaSmallerThanDp}

Here, we derive the diffusion coefficient for an isolated SPR subject to fluctuations. From the equations of motion Eq.~(1) and Eq.~(2), it is clear that the temporal evolution 
of an isolated SPR can be expressed as:
\begin{eqnarray}
\nonumber
\dot{\mathbf{x}}_i &=& v_0  \mathbf{V}(\theta_i) + \tilde{\sigma}_{i\parallel}(t)  \mathbf{V}(\theta_i)  + \tilde{\sigma}_{i\perp}(t)  \mathbf{V}_{\perp}(\theta_i) \\
\label{eq:evol_x_si}
 &=& \mathbf{u}(t)   \\
\label{eq:evol_theta_si}
\dot{\theta}_i       &=&      \tilde{\xi}_{i}(t) \, , 
\end{eqnarray}
where in Eq.~(\ref{eq:evol_x_si}) we have expressed explicitly  $\dot{\mathbf{x}}_i$ in terms of the components parallel, $\mathbf{V}(\theta_i)$, and perpendicular, 
$\mathbf{V}_{\perp}(\theta_i)$,  to the long axis of the rod. The terms $\tilde{\sigma}_{i\parallel}(t)$, $\tilde{\sigma}_{i\perp}(t)$, and $\tilde{\xi}_{i}(t)$ refer to 
independent delta-correlated noises~\cite{gardiner}, as in Eq.~(1) and Eq.~(2), but where we have absorbed the corresponding friction coefficients into the noise definition. 
The term $\mathbf{u}(t)$ in Eq.~(\ref{eq:evol_x_si}) refers to the instantaneous velocity of the particle and should not be confused with the interaction potential. 
The position of the SPR at time $t$ can be formally expressed  as $\mathbf{x}_i = \int_0^{t} ds \mathbf{u}(s)$, and  the mean-square displacement, through the Taylor-Kubo formula~\cite{kubo1957}, is given by:
\begin{eqnarray}
\label{eq:MSD}
\langle \mathbf{x}_i^2(t) \rangle = \int_0^t ds \int_0^t ds' \langle \mathbf{u}(s) . \mathbf{u}(s') \rangle  \, . 
\end{eqnarray}
To compute $\langle \mathbf{u}(s) . \mathbf{u}(s') \rangle$ and finally to evaluate the integral, we use that $\langle \tilde{\sigma}_{i\parallel}(t) \rangle = \langle \tilde{\sigma}_{i\perp}(t) \rangle = \langle  \tilde{\xi}_{i}(t) \rangle =0$ and the autocorrelations of the independent delta-correlated noises~\cite{gardiner} that read: 
\begin{eqnarray}
\langle  \tilde{\sigma}_{i\parallel}(s)   \tilde{\sigma}_{i\parallel}(s')  \rangle &=& 2\tilde{D}_{\parallel} \delta(s-s') \\
\langle  \tilde{\sigma}_{i\perp}(s)   \tilde{\sigma}_{i\perp}(s)  \rangle &=& 2\tilde{D}_{\perp} \delta(s-s') \\
\langle  \tilde{\xi}_{i}(s) \tilde{\xi}_{i}(s')    \rangle &=& 2\tilde{D}_{\theta} \delta(s-s') \, , 
\end{eqnarray}
while other combinations vanish, i.e., $\langle  \tilde{\sigma}_{i\parallel}(s)   \tilde{\sigma}_{i\perp}(s')  \rangle =  \langle  \tilde{\sigma}_{i\parallel}(s)  \tilde{\xi}_{i}(s')     \rangle =  \langle  \tilde{\sigma}_{i\perp}(s)  \tilde{\xi}_{i}(s')     \rangle = 0$. 
This correlations allow us to compute  $\langle \mathbf{x}_i^2(t) \rangle$, which  for $t\gg 1/\tilde{D}_{\theta}$ takes the form:
\begin{eqnarray}
\langle \mathbf{x}_i^2(t) \rangle =  2 \frac{v_0^2}{\tilde{D}_{\theta}} t + 2 (\tilde{D}_{\parallel} + \tilde{D}_{\perp}) t \,.
\end{eqnarray}
The diffusion coefficient is defined in two dimensions as $D = \lim_{t \to \infty}  \langle \mathbf{x}_i^2(t) \rangle /(4t)$, and thus we obtain: 
\begin{eqnarray}
D  =   \frac{v_0^2}{2 \tilde{D}_{\theta}}  + \frac{\tilde{D}_{\parallel} + \tilde{D}_{\perp}}{2}  = D_a + D_p\,, 
\end{eqnarray}
where we define the active diffusion coefficient by $D_a= \frac{v_0^2}{2 \tilde{D}_{\theta}}$ and the passive diffusion coefficient by $D_p = \frac{\tilde{D}_{\parallel} + \tilde{D}_{\perp}}{2}$. 
Notice that if the noise terms are of thermal origin in~Eqs.~(\ref{eq:evol_x_si}) and~(\ref{eq:evol_theta_si}), then $\tilde{D}_{\parallel}  \propto T$, $\tilde{D}_{\perp}  \propto T$, 
and $\tilde{D}_{\theta}  \propto T$, where $T$ is the temperature. The ratio between any of these coefficient, e.g. $\tilde{D}_{\parallel}/\tilde{D}_{\perp}$ is proportional to the 
ratio between the corresponding drag coefficients: as example $\tilde{D}_{\parallel}/\tilde{D}_{\perp}=\zeta_{\perp}/\zeta_{\parallel}$. 
From this, we learn that while $D_p \propto T$, $D_a \propto 1/T$. 
Finally, we notice that we can express, for the reasons given above, $\tilde{D}_{\parallel} = \tilde{D}_{\theta} \zeta_{\theta}/\zeta_{\parallel}$ and $\tilde{D}_{\perp} = \tilde{D}_{\theta} \zeta_{\theta}/\zeta_{\perp}$. 
 For the values of $D_\theta$, $v_0$, $\zeta_{\parallel}$, $\zeta_{\perp}$ and $\zeta_{\theta}$ used in the main text, the ratio between the active and passive diffusion coefficient is such  that $D_p/D_a \sim 10^{-3}$, i.e.  $D_p\ll D_a$. 
 In such a regime, we can ignore the contribution of $D_p$, meaning that we can neglect  $\tilde{\sigma}_{i\parallel}$ and  $\tilde{\sigma}_{i\perp}$, and focus exclusively on the role of $\tilde{\xi}_{i}$ on $D_a$. 

\section{Exploration of the parameter space}
\label{phase-diagrams}

\begin{figure}
\begin{center}
\resizebox{\columnwidth}{!} {\includegraphics{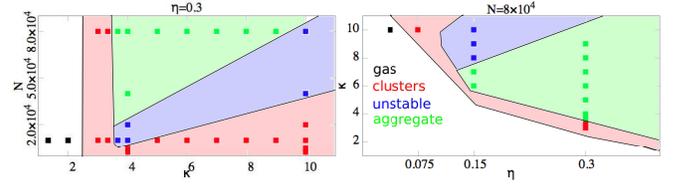}}
\caption{Exploration of the parameter space (phase diagram). Each squared point in the plots correspond to a parameter set $\{N, \eta, \kappa\}$ for which we have performed simulations. The color of a point encodes the observed state of active matter. Black: disordered, not phase-separated state (gas). Red: ordered phase-separated state (polar clusters or polar band). Green: disordered phase-separated state (aggregate). Blue: unstable situation, where the system oscillates between the ordered phase-separated state and the disordered phase-separated state. The color code for the regions of the phase diagram is the same as for the simulation points.
}
\label{fig-phase-diagrams}
\end{center}
\end{figure}

The left-hand panel of  Fig.~\ref{fig-phase-diagrams} indicates, the parameter sets  $\{N, \kappa\}$  for which we have performed simulations at packing fraction $\eta=0.3$, and the observed states of active matter. Note that when increasing $N$, the orientationally ordered phase ({\it i.e.} clusters) progressively disappears and is replaced by an disordered phase (aggregate).

The right-hand panel of Fig.~\ref{fig-phase-diagrams} corresponds to the phase diagram for large systems ($N=80000$). We observe that the area of the phase space corresponding to orientational order (clusters) is reduced to a small band (which should completely disappear when further increasing $N$).

% \begin{figure}[h!]

\section{Polar order and clusters}
\label{sec:polar-order-and-clusters}

In Figs.~1 and 3 of the paper we have investigated the global polar order $S_1$ in the system. We observed polar order for $\kappa>\kappa_c$ in Fig.~1 and for $N<N_*$ in Fig.~3. If there are only few giant polar clusters (or even a single percolating band) in the system, the observed global polar order directly results from the polar order within these few polar clusters. However, if there are many clusters, their orientations may be correlated, which also contributes to global polar order. For very big systems ({\it i.e.} in the thermodynamic limit), global polar order can only be observed if the orientations of the clusters are correlated. Here we attempt a quantification of the impact of cluster correlations on global polar order. Therefore, we assume that polar order within clusters only depends on cluster size $m$ and denote it as $S_{1,m}= 1/n_m \sum^{n_m}_{h=1} |1/m \sum^{m}_{k=1} e^{\iota \theta_k}|$: the polar order associated to the cluster of mass $m$. We then measure the cluster size distribution $p(m)$. This permits us to compute the value of the global order polar parameter $S_{1,uncorrelated}$ under the hypothesis that there are no correlations, i.e. we assume that, at a given time $t$, any cluster $j$ has a moving direction $\hat{\theta}_j(t)$ (defined as the average moving direction of all the rods within the cluster) which is distributed according to a uniform random distribution.

%\begin{figure*}[htbp]
%\begin{figure}[h!]
\begin{figure}
\begin{center}
\resizebox{\columnwidth}{!} {\includegraphics{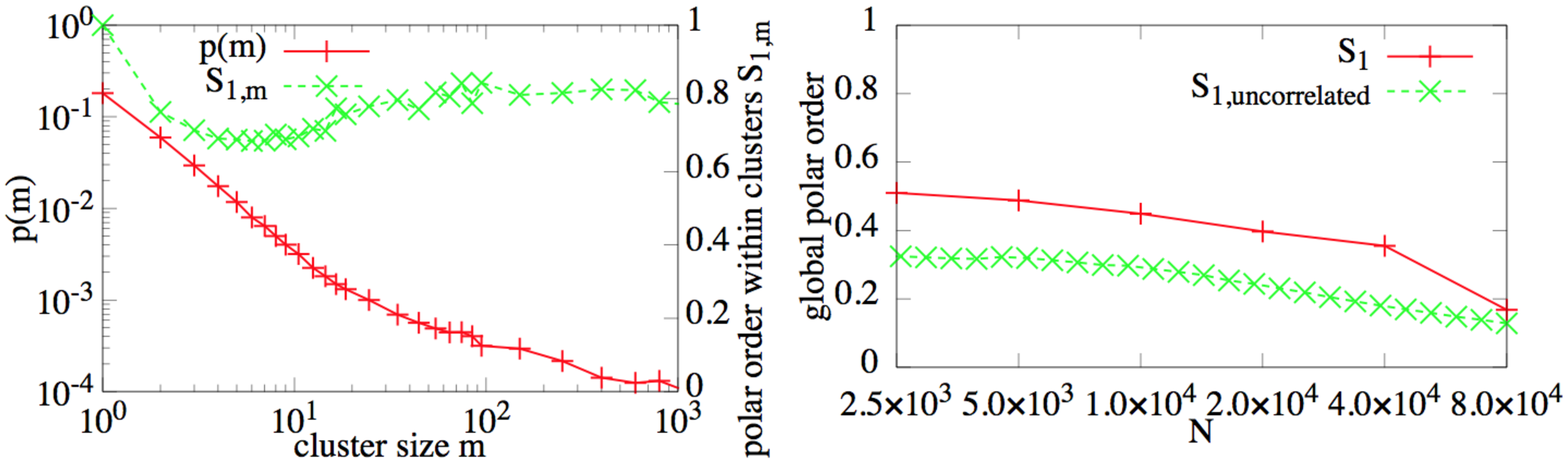}}
\caption{
Left: cluster size distributions $p(m)$ and the corresponding polar-order-within-clusters-distributions $S_{1,m}$ (i.e. the polar order within clusters as function of their size $m$). Right: Global polar order parameter effectively observed $S_1$ and global polar order parameter obtained under the hypothesis that cluster orientations are fully decorrelated $S_{1,uncorrelated}$. Simulations correspond to $\eta=0.15$, $\kappa=10$ (and $N=20000$  for the $S_{1,m}$-distribution, but this distribution is very similar for the other values of $N$).
%Top panel : $\eta=0.075$, $\kappa=10$. Bottom panel: $\eta=0.15$, $\kappa=10$. %J'AI ENLEVE LA FIGURE ETA=0.3 CAR MON ALGO NE FAIT PLUS AUCUN SENS LORSQU'IL Y A DES CLUSTERS DE GRANDE TAILLE (COMPARE A LA TAILLE DU SYSTEME N).
}
\label{fig-cluster-correlations}
\end{center}
\end{figure}

The mathematical derivation is the following: Starting from the definition of the global polar order parameter given by Eq.~3 of the main text, we rewrite the average over all particles as an average over all clusters, i.e. 

\begin{eqnarray}
\nonumber
S_{1,uncorrelated}
&=& \langle | \frac{1}{N}\sum_{i=1}^N e^{\iota\theta_i(t)} | \rangle_t\\
\nonumber
&=& \langle | \frac{1}{N}\sum_{j=1}^{N_c(t)}\sum_{i=1}^{{m}_j(t)} e^{\iota\theta_i(t)} | \rangle_t\\
% &\sim& \langle | \frac{1}{N}\sum_{j=1}^{N_c(t)}{m}_j(t) {S}_{1,m}e^{\iota\theta_j(t)} | \rangle_t\\
% &\sim&  \langle | \frac{1}{N}\sum_{m=1}^{N} m {S}_{1,m} \sum_{j=1}^{{n}_m} e^{\iota \hat{\theta}_j} |\rangle_{\{\hat{\theta}_j\}} \, ,
&\sim& \langle | \frac{1}{N}\sum_{j}{m}_j {S}_{1,m}e^{\iota \hat{\theta}_j} | \rangle_{ \{ m_j, \hat{\theta}_j \}}
%FAUX &\sim&  | \frac{1}{N}\sum_{m=1}^{N} {S}_{1,m} \sum_{j=1}^{{n}_m} e^{\iota \hat{\theta}_j(t)} | ,\\
%&=& \langle | \langle \exp(\iota \theta_i(t)) \rangle_i | \rangle_t \\
%&=& \langle | \langle \hat{m}_j(t) \exp(\iota \hat{\theta}_j(t)) \rangle_j | \rangle_t \\
%&=& \langle | \langle \hat{m}_j(t) \hat{S}_{1,j}(t) \exp(\iota \hat{\theta}_j(t)) \rangle_j | \rangle_t, \\
%&=& \langle  | \langle w_j \rangle_j | \rangle_t, \\
\end{eqnarray}

where $m_j$ is taken from the steady state cluster size distribution $p(m)$ and $\hat{\theta}_j$ from an homogeneous distribution between $[0, 2\pi]$, such that the number of rods is $N$. 
% The exact calculation is approximated by  a sum over the number of clusters, the cluster sizes being drawn randomly from the cluster size distribution $p(m)$ under the constraint that the total number of rods is $N$. 
This is an approximation because we replace the polar order of a cluster by the average order of a cluster of the corresponding mass, $S_{1,m}$.
The distributions $p(m)$ and $S_{1,m}$ have been directly measured from simulations, see left and central plot of Fig.~\ref{fig-cluster-correlations}.
The final step is to assume that each 
cluster points in a (uniform) random direction $\hat{\theta}_j(t)$ as explained above. The computation of the global order parameter is done by performing a Monte Carlo algorithm, required to average over the random direction of the cluster (and the random cluster size). We refer to this quantity as  $S_{1,uncorrelated}$, see  Fig.~\ref{fig-cluster-correlations}.
We then compare $S_{1,uncorrelated}$ to the effectively observed polar order $S_1$ (see the right-hand plot in Fig.~\ref{fig-cluster-correlations}). For all investigated packing fraction $\eta$  (but in particular for high packing fractions) and system sizes $N$, the observed polar order ($S_1$) is systematically bigger than the one resulting from the cluster decorrelation hypothesis ($S_{1,uncorrelated}$). This indicates the presence of cluster correlations. Nevertheless, we observe that $S_1$ becomes closer to $S_{1,uncorrelated}$ when the system size $N$ increases. This suggests that cluster correlations are of finite size (i.e. the orientations of clusters which are sufficiently far away from each other are no more correlated). This is a strong indication that in the thermodynamical, limit cluster-cluster correlations are weak and cannot produce (global) polar order.

\bibliographystyle{apsrev}

%% \bibliography{biblio_spp_spr}

\end{document}